\documentclass[a4paper]{jpconf}

\usepackage{graphicx}
\usepackage{amsmath}
\usepackage{amssymb}
\usepackage{graphicx}
\usepackage{color}
\usepackage{hyperref}
\usepackage{xspace}

\DeclareFontFamily{OT1}{pzc}{}
\DeclareFontShape{OT1}{pzc}{m}{it}%
{<-> s * [1.15] pzcmi7t}{}
\DeclareMathAlphabet{\mathpzc}{OT1}{pzc}{m}{it}

\begin{document}
\title{DSE inspired model for the pion's valence dressed-quark GPD}

\author{L. Chang$^1$, C. Mezrag$^2$, H. Moutarde$^2$, C. D. Roberts$^3$, \\ J. Rodr\'{\i}guez-Quintero$^4$, F. Sabati\'e$^2$}

\address{$^1$ CSSM, School of Chemistry and Physics
University of Adelaide, Adelaide SA 5005, Australia}

\address{$^2$ Centre de Saclay, IRFU/Service de Physique Nucl\'eaire, F-91191 Gif-sur-Yvette, France}

\address{$^3$ Physics Division, Argonne National Laboratory, Argonne, Illinois 60439, USA}

\address{$^4$ Departamento de F\'isica Aplicada, Facultad de Ciencias Experimentales, Universidad de Huelva, Huelva E-21071, Spain}


\begin{abstract}
We sketch here an approach to the computation of genaralised parton distributions (GPDs), based upon a rainbow-ladder (RL) truncation of QCD's Dyson-Schwinger equations and exemplified via the pion's valence dressed-quark GPD, $H_\pi^{\rm v}(x,\xi,t)$.  Our analysis focuses on the case of zero skewness, $\xi=0$, and underlines that the impulse-approximation used hitherto to define the pion's valence dressed-quark GPD is generally invalid owing to omission of contributions from the gluons which bind dressed-quarks into the pion.  A simple correction enables us to identify a practicable improvement to the approximation for $H_\pi^{\rm v}(x,0,t)$, expressed as the Radon transform of a single amplitude. Therewith we obtain results for $H_\pi^{\rm v}(x,0,t)$ and the associated impact-parameter dependent distribution, $q_\pi^{\rm v}(x,|\vec{b}_\perp|)$, which provide a qualitatively sound picture of the pion's dressed-quark structure at an hadronic scale.  
\end{abstract}

\section{Introduction}

The current contribution is aimed at reporting on the results, recently presented in Refs.\,\cite{Chang:2014lva,Mezrag:2014jka}, concerning the computation of the pion's valence dressed-quark GPD, based upon a rainbow-ladder (RL) truncation of QCD's Dyson-Schwinger equations (DSE), and its connection to the parton distribution function (PDF) in the appropriate forward limit. 


After more than forty years of studying PDFs, since quarks were discovered in a series of deep inelastic scattering (DIS) experiments at the Stanford Linear Accelerator Center \cite{Friedman:1991nq,Taylor:1991ew,Kendall:1991np}, a good deal is known about hadron light-front structure. Notwithstanding that, much more needs to be understood, particularly, e.g., in connection with the distribution of helicity \cite{Myhrer:2009uq,Jimenez-Delgado:2013sma}.  Moreover, PDFs only describe hadron light-front structure incompletely because inclusive DIS measurements do not yield information about the distribution of partons in the plane perpendicular to the bound-state's total momentum; i.e., within the light front.  Such information is expressed in generalised parton distributions (GPDs) \cite{Dittes:1988xz,Mueller:1998fv,Radyushkin:1996nd,Ji:1996nm}, which are accessible via deeply virtual Compton scattering on a target hadron, $T$; viz., $\gamma^\ast(q) T(p) \to \gamma^\ast(q^\prime)T(p^\prime)$, so long as at least one of the photons [$\gamma^\ast(q)$, $\gamma^\ast(q^\prime)$] possesses large virtuality, and in the analogous process of deeply virtual meson production: $\gamma^\ast(q) T(p) \to M(q^\prime)T(p^\prime)$.  Importantly [see Sect. 2], GPDs connect PDFs with hadron form factors because any PDF may be recovered as a forward limit of the relevant GPD and any hadron elastic form factors can be expressed via a GPD-based sum rule.  The potential that GPDs hold for providing manifold insights into hadron structure has led to intense experimental and theoretical activity \cite{Burkardt:2002hr,Diehl:2003ny,Belitsky:2005qn,Boffi:2007yc,Guidal:2013rya}.

Most of the constraints that apply to GPDs are fulfilled when the GPD is written as a double distribution
\cite{Mueller:1998fv,Radyushkin:1998es,Radyushkin:1998bz}, which is equivalent to expressing the GPD as a Radon transform \cite{Teryaev:2001qm}:
\begin{equation}
\label{radon}
H(x,\xi,t) = \int_{|\alpha|+|\beta|\leq 1} \rule{-5ex}{0ex} d\alpha \, d\beta \,
\delta(x-\alpha-\xi\beta)\, [F(\alpha,\beta,t) + \xi\, G(\alpha,\beta,t)]\,,
\end{equation}
where the variables $x$, $\xi$, $t$ are defined following Eq.~\eqref{eq-def-GPD-H-spinless-target} and, at leading-twist, $F$, $G$ have operator definitions analogous to the GPD itself.  In order to obtain insights into the nature of hadron GPDs, it has been common to model the Radon amplitudes, $F$, $G$, following Refs.\,\cite{Musatov:1999xp}.  This approach has achieved some phenomenological success \cite{Guidal:2013rya,Mezrag:2013mya}; but more flexible parametrisations enable a better fit to data \cite{Kumericki:2008di}.  Such fits played a valuable role in establishing the GPD framework;
but if one wishes to use measured GPDs as a means by which to validate our basic perception of strong interactions in the Standard Model, then data fitting is inadequate.  Instead, it is necessary to compute GPDs using a framework that possesses a direct connection with QCD.  This observation is highlighted by experience drawn from the simpler case of the pion's valence-quark PDF \cite{Chang:2014lva}.  Herein, we therefore adopt a different approach, sketching a procedure for the computation of hadron GPDs based on the example provided by the pion's valence-quark PDF.

\section{General features of pion GPDs}

{F}rom a quark model perspective, in the isospin symmetric limit, the pion is a  quantum mechanical bound-state of two equal-mass constituents and it is therefore the simplest hadronic bound-state.  That is a misapprehension, however.  Owing to the connection between pion properties and dynamical chiral symmetry breaking (DCSB); i.e., its dichotomous nature as a Goldstone mode and relativistic bound-state \cite{Maris:1997hd,Qin:2014vya}, a veracious description of the pion is only possible within a framework that faithfully expresses symmetries and their breaking patterns.  The Dyson-Schwinger equations (DSEs) fulfill this requirement \cite{Chang:2011vu,Bashir:2012fs,Cloet:2013jya} and hence we employ that framework to compute pion properties herein.

Notwithstanding the complex nature of the pion bound-state, it is still a $J=0$ system and hence for a vector probe there is only one GPD associated with a quark $q$ in the pion ($\pi^\pm$, $\pi^0$).  It is defined by the matrix element
\begin{eqnarray}
\label{eq-def-GPD-H-spinless-target}
H^q_{\pi}( x, \xi, t ) =  \int \frac{\mathrm{d}^4z}{4\pi} \, e^{i x P\cdot z}\,
\delta(n\cdot z) \, \delta^2(z_\perp) \,\langle\pi(P_+)| \bar{q}\left(-z/2\right)n\cdot \gamma \; 
q\left(z/2\right) |\pi(P_-)\rangle,
\end{eqnarray}
where: $k$, $n$ are light-like four-vectors, satisfying $k^2=0=n^2$, $k\cdot n=1$; $z_\perp$ represents that two-component part of $z$ annihilated by both $k$, $n$; and $P_\pm = P \pm \Delta/2$.
%
In Eq.\,\eqref{eq-def-GPD-H-spinless-target}, $\xi = -n\cdot \Delta/[2 n\cdot P]$ is the ``skewness'', $t=-\Delta^2$ is the momentum transfer, and $P^2 = t/4-m_\pi^2$, $P\cdot \Delta=0$.
The GPD also depends on the resolving scale, $\zeta$.  Within the domain upon which perturbation theory is valid, evolution to another scale $\zeta^\prime$ is described by the ERBL equations \cite{Efremov:1979qk,Lepage:1980fj} for $|x|<\xi$ and the DGLAP equations \cite{Dokshitzer:1977,Gribov:1972,Lipatov:1974qm,Altarelli:1977} for $|x|>\xi$, where $\xi \geq 0$.

In order to produce quantities that are gauge invariant for all values of $z$, Eq.\,\eqref{eq-def-GPD-H-spinless-target} should contain a Wilson line, ${\cal W}[-z/2,z/2]$, between the quark fields.  Notably, for any light-front trajectory, ${\cal W}[-z/2,z/2]\equiv 1$ in lightcone gauge: $n\cdot A=0$, and hence the Wilson line does not contribute in this case.  On the other hand, light-cone gauge is seldom practicable in either model calculations or quantitative nonperturbative analyses in continuum QCD.  Indeed, herein, as typical of nonperturbative DSE studies, we employ Landau gauge because, \emph{inter alia} \cite{Bashir:2008fk,Bashir:2009fv}: it is a fixed point of the renormalisation group; and a covariant gauge, which is readily implemented in numerical simulations of lattice-QCD.  It is therefore significant that ${\cal W}[-z/2,z/2]$ is not quantitatively important in the calculation of the leading-twist contributions to numerous matrix elements \cite{Kopeliovich:2011rv}.

It is worth recapitulating here upon some general properties of GPDs.  Most generally, Poincar\'e covariance entails that GPDs are only nonzero on $x\in (-1,1)$.  Moreover, owing to time-reversal invariance, $H^{q}(x,\xi,t) =H^{q}(x,-\xi,t)$.  Kinematically, the skewness is bounded: $\xi\in[-1,1]$, but $\xi\in[0,1]$ for all known processes that provide empirical access to GPDs.

Focusing on the pion, a charge conjugation mapping between charged states entails $H^{u,d}_{\pi^+}(x,\xi,t) =  - H^{u,d}_{\pi^-}(-x,\xi,t)$; and consequently, in the isospin symmetric limit:
\begin{equation}
H^u_{\pi^+}(x,\xi,t) = - H^d_{\pi^+}(-x,\xi,t).
\label{eq:G-parity-GPD}
\end{equation}
It follows that the isospin projections:
\begin{equation}
H^{I}(x,\xi,t):=
H_{\pi^+}^u(x,\xi,t) + (-1)^I H_{\pi^+}^d(x,\xi,t), I=0,1\,, \label{eq:H_isoscalar_both}
\end{equation}
have well-defined symmetry properties under $x\leftrightarrow -x$; viz.,
$H^{0}$ is odd and $H^{1}$ is even.

Returning to the definition in Eq.\,\eqref{eq-def-GPD-H-spinless-target}, it is plain that if one considers the forward limit: $\xi=0$, $t=0$, then $x$ is Bjorken-$x$ and the GPD reduces to a PDF; viz.,
\begin{equation}
H_\pi^q(x,0,0) = \left\{
\begin{array}{rr}
q^\pi(x), & x>0\\
-\bar q^\pi(-x), & x<0
\end{array}\right.\,.
\label{PDFconnection}
\end{equation}
Moreover, irrespective of the value of $\xi$, the electromagnetic pion form factor may be computed as
\begin{align}
F_{\pi^+}(\Delta^2) &= \int_{-1}^1 dx\, [ e_u  H_{\pi^+}^u(x,\xi,-\Delta^2) + e_d  H_{\pi^+}^d(x,\xi,-\Delta^2) ]\quad\label{Fpiconnection1}\\
&=: e_u F_{\pi^+}^u(\Delta^2) + e_d F_{\pi^+}^d(\Delta^2)
= F_{\pi^+}^u(\Delta^2)\,, \label{Fpiconnection2}
\end{align}
where $e_{u,d}$ are the quark electric charges in units of the positron charge and we have used Eq.\,\eqref{eq:G-parity-GPD} to show $F_{\pi^+}^d(\Delta^2) = -  F_{\pi^+}^u(\Delta^2)$.  Additional information may be found elsewhere \cite{Mezrag:2014tva}.


\section{Heuristic Example}

Imagine a bound-state of two scalar particles with effective mass $\sigma$ and suppose that the interaction between them is such that it produces a light-front wave function of the form ($\bar x=1-x$):
\begin{equation}
\label{psimodel}
\psi(x,k_\perp^2) = \sqrt{\frac{15}{2\pi \,\sigma^2 }}\,\frac{\sqrt{x \bar x}}{1+k_\perp^2/(4 \,\sigma^2 x\bar x)}\theta(x)\theta(\bar x)\,.
\end{equation}
(A merit of considering a bound-state of scalar constituents is that in describing the wave function of the composite system one avoids the complication of Melosh rotations, which arise in treating spin states in light-front quantum mechanics \cite{Coester:1992cg}.)  If the skewness is zero, in which case the momentum transfer is purely light-front transverse, then the GPD for this system can be written as a wave function overlap \cite{Burkardt:2000za,Diehl:2000xz,Burkardt:2002hr,Diehl:2003ny}:
\begin{equation}
\label{overlapGPD}
H_\sigma(x,0,-\Delta_\perp^2) = \int d^2 k_\perp \, \psi(x,k_\perp+(1-x) \Delta_\perp) \, \psi(x,k_\perp)\,.
\end{equation}
This entails
\begin{equation}
\label{HPositive}
\{ H_\sigma(x,0,\Delta_\perp^2)>0 : x\in[-1,1], \Delta_\perp^2\geq 0 \}\,.
\end{equation}

Owing to the simplicity of the starting point, Eqs.\,\eqref{psimodel} and \eqref{overlapGPD} allow one to obtain an algebraic expression for the GPD; viz., with $z^2 =\Delta_\perp^2 (1-x)/4x\sigma^2 $, then
\begin{align}
&H_\sigma(x,0,-\Delta_\perp^2)  = 30 (1-x)^2 x^2
{\cal C}(z)\,\theta(x)\theta(\bar x)\,,
\\
&{\cal C}(z) = \frac{\ln\left[ \left(z^3+\left(z^2+1\right)
   \sqrt{z^2+4}+3 z\right)/
   \left(\sqrt{z^2+4}-z\right)\right]}{z \sqrt{z^2+4}}\,.
   \label{eqCz}
\end{align}
Some further analysis reveals that $C(z)$ decreases monotonically away from its maximum value ${\cal C}(z=0)=1$.  Consequently, $H_\sigma(x,0,0)= 30 (1-x)^2 x^2\theta(x)\theta(\bar x) $, which is an excellent approximation to the pion's valence dressed-quark PDF \cite{Chang:2014lva}; and whilst the maximum of $H_\sigma(x,0,-\Delta^2)$ lies at $x=1/2$ for $\Delta^2=0$, this peak shifts to $x=1$ with increasing $\Delta^2$, consistent with an expectation that for $\Delta^2\gg \sigma^2$ the interaction probability is largest when the probe and hadron are collinear \cite{Burkardt:2002hr}.

The Hankel transform:
\begin{equation}
\label{defqxb}
q_\sigma(x,|\vec{b}_\perp|) =
\int \frac{d|\Delta_\perp|}{2\pi}|\Delta_\perp| J_0(|\vec{b}_\perp||\vec{\Delta}_\perp|) H(x,0,-\Delta_\perp^2)\,,
\end{equation}
defines the system's impact-parameter-dependent (IPD) GPD \cite{Burkardt:2000za}. It is a density that describes the probability of finding a parton within the light-front at a transverse position $\vec{b}_\perp$ from the hadron's centre of transverse momentum (CoTM).  Since $H(x,0,-\Delta_\perp^2)$ is a positive-definite, monotonically decreasing function of $\Delta_\perp^2$ for each $x$, the global maximum of $q_\sigma(x,|\vec{b}_\perp|)$ is located at $|\vec{b}_\perp|=0$ and $q_\sigma(x,|\vec{b}_\perp|)$ is a monotonically decreasing, positive-definite function away from that maximum.

The value of $x$ at which the global maximum in $q_\sigma(x,|\vec{b}_\perp|)$ occurs is determined by the system's dynamics.  Considering the hadron's valence dressed-parton structure, one extreme is achieved if ${\cal C}(z)$ is independent of $x$: the maximum of $q_\sigma(x,|\vec{b}_\perp|)$ is then located at $(x=1/2,|\vec{b}_\perp|=0)$.  In realistic cases, the necessary $(x,\Delta_\perp^2)$ correlations in ${\cal C}(z)$ and this function's general properties act to shift the maximum to $x>1/2$.  Using Eqs.\,\eqref{psimodel} and \eqref{overlapGPD}, $q_\sigma(x,|\vec{b}_\perp|)$ peaks at $(x=0.72,|\vec{b}_\perp|=0)$.  One may also consider the path followed by the maximum as one increases $|\vec{b}_\perp|$ away from zero.  To that end, observe from Eq.\,\eqref{defqxb} that for $|\vec{b}_\perp|\gg 1/\sigma$ the $x$-dependence of $q_\sigma(x,|\vec{b}_\perp|)$ is dominated by $H(x,0,-\Delta_\perp^2 \simeq 0)$, which peaks at $x=1/2$.  The nature of ${\cal C}(z)$ then entails that the peak in the valence dressed-quark IPD GPD drifts monotonically toward $x=1/2$ as $\sigma |\vec{b}_\perp|\to 0$.


\section{Pion's valence dressed-quark GPD}
In order to compute an approximation to the valence-quark piece of the GPD expressed in Eq.\,\eqref{eq-def-GPD-H-spinless-target} we adapt the method used successfully elsewhere to compute the pion's valence-quark distribution function \cite{Chang:2014lva} and elastic form factor \cite{Chang:2013nia}.  Consider, therefore,
\begin{equation}
2 H_{\pi}^{\rm v}(x,\xi,t) =
 N_c \mathrm{tr}\rule{-0.5ex}{0ex}
 \int_{d\ell}\,\delta_n^{xP}(\ell)\,i{\Gamma}_\pi(\ell_+^{\rm R};-P_+ )
S(\ell_+) \,in\cdot\Gamma(\ell_+,\ell_-) \, S(\ell_-) i\Gamma_\pi(\ell_-^{\rm R}; P_- )\,,
\label{eq:TriangleDiagrams}
\end{equation}
where $\int_{d\ell} := \int \frac{d^4\ell}{(2\pi)^4}$ is a translationally invariant regularisation of the integral; $\delta_n^{xP}(\ell):= \delta(n\cdot \ell - x n\cdot P)$; the trace is over spinor indices; $\eta\in[0,1]$, $\bar\eta=1-\eta$; $\ell_+^{\rm R}=\bar\eta\ell_+ +\eta\ell_P$,
$\ell_-^{\rm R}=\eta\ell_- +\bar\eta\ell_P$,
$\ell_\pm = \ell \pm \Delta/2$, $\ell_P=\ell-P$.  (N.B.\ Owing to Poincar\'e covariance, no observable can legitimately depend on $\eta$; i.e., the definition of the relative momentum.)
So long as each of the dressed-quark propagators, $S(\ell)$, on the right-hand-side (rhs) of Eq.\,\eqref{eq:TriangleDiagrams} is computed using the rainbow truncation of QCD's gap equation, and both the pion Bethe-Salpeter amplitudes, $\Gamma_\pi(\ell;P)$, and the dressed-quark-photon vertex, $\Gamma_\mu(\ell_f,\ell_i)$, are calculated in the associated ladder truncation of the relevant Bethe-Salpeter equations then $H_{\pi}(x,\xi,t)$, thus computed and inserted in Eq.\,\eqref{Fpiconnection1}, provides the leading-order contribution to the pion's electromagnetic form factor in the most widely used, symmetry preserving truncation of QCD's DSEs: the rainbow-ladder (RL) truncation \cite{Munczek:1994zz,Bender:1996bb}, whose strengths and limitations are detailed elsewhere \cite{Chang:2011vu,Bashir:2012fs,Cloet:2013jya}.

Given its connection with a reliable scheme for computing $F_\pi(Q^2)$, it was long thought \cite{Hecht:2000xa,Nguyen:2011jy} that Eq.\,\eqref{eq:TriangleDiagrams} would also be an adequate starting point for computation of the pion's valence-quark PDF, $q_V^\pi(x)$.
However, as explained in Ref.\,\cite{Chang:2014lva}, that is not true.

\begin{figure}[h]
\begin{minipage}{38pc}
\includegraphics[width=11.5pc]{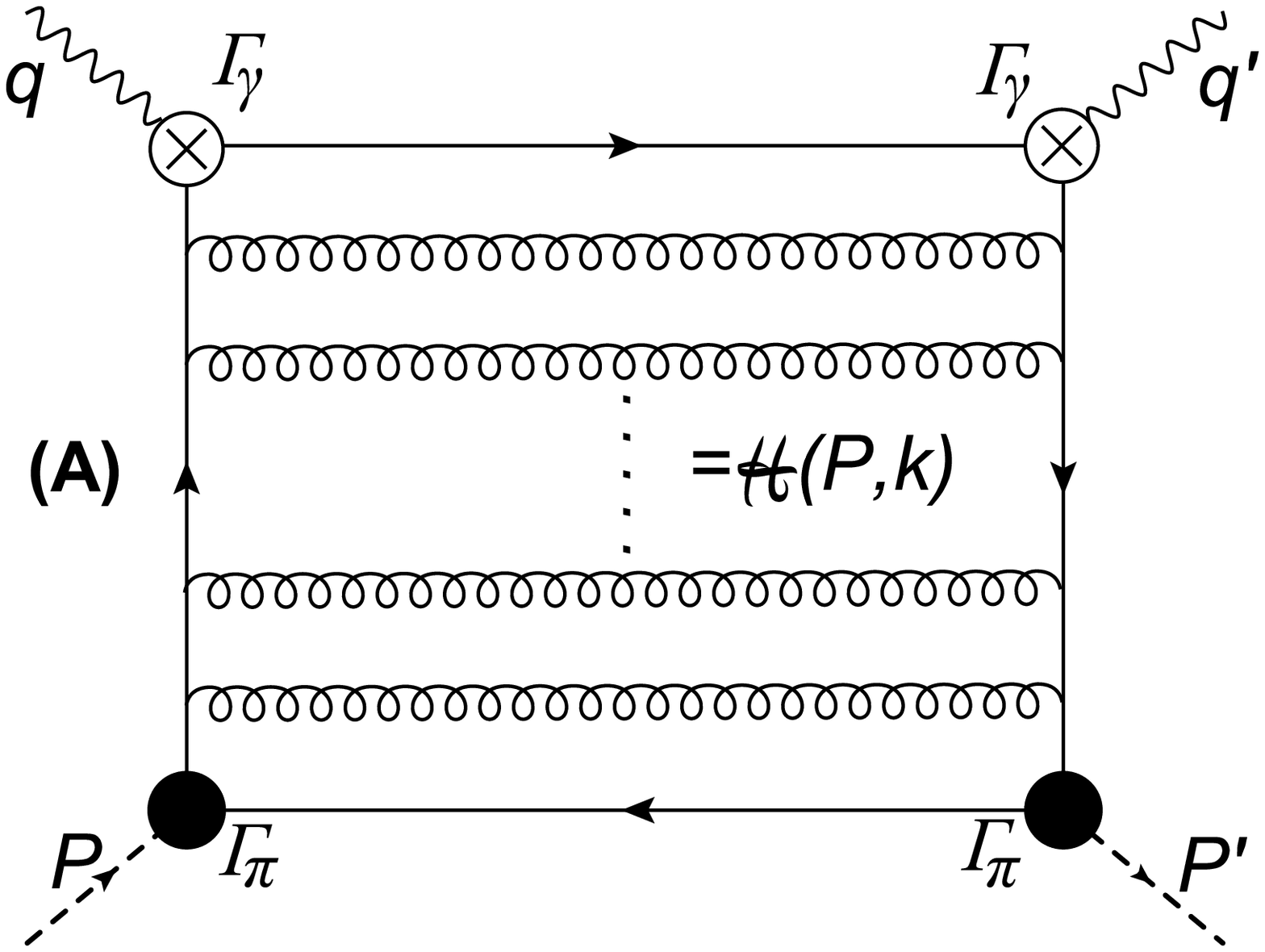} \hspace{0.35pc}
\includegraphics[width=11.5pc]{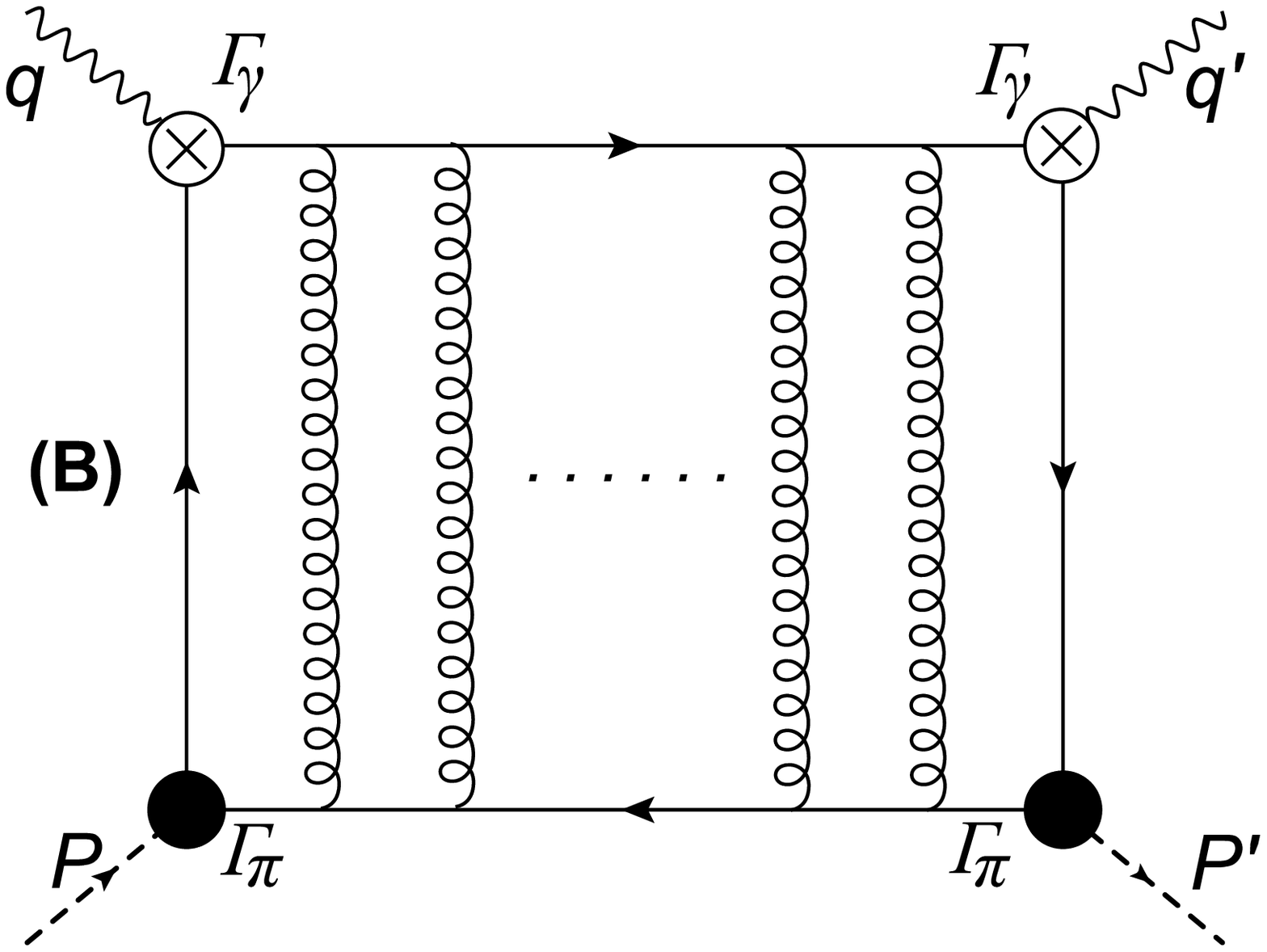} \hspace{0.35pc}
\includegraphics[width=11.5pc]{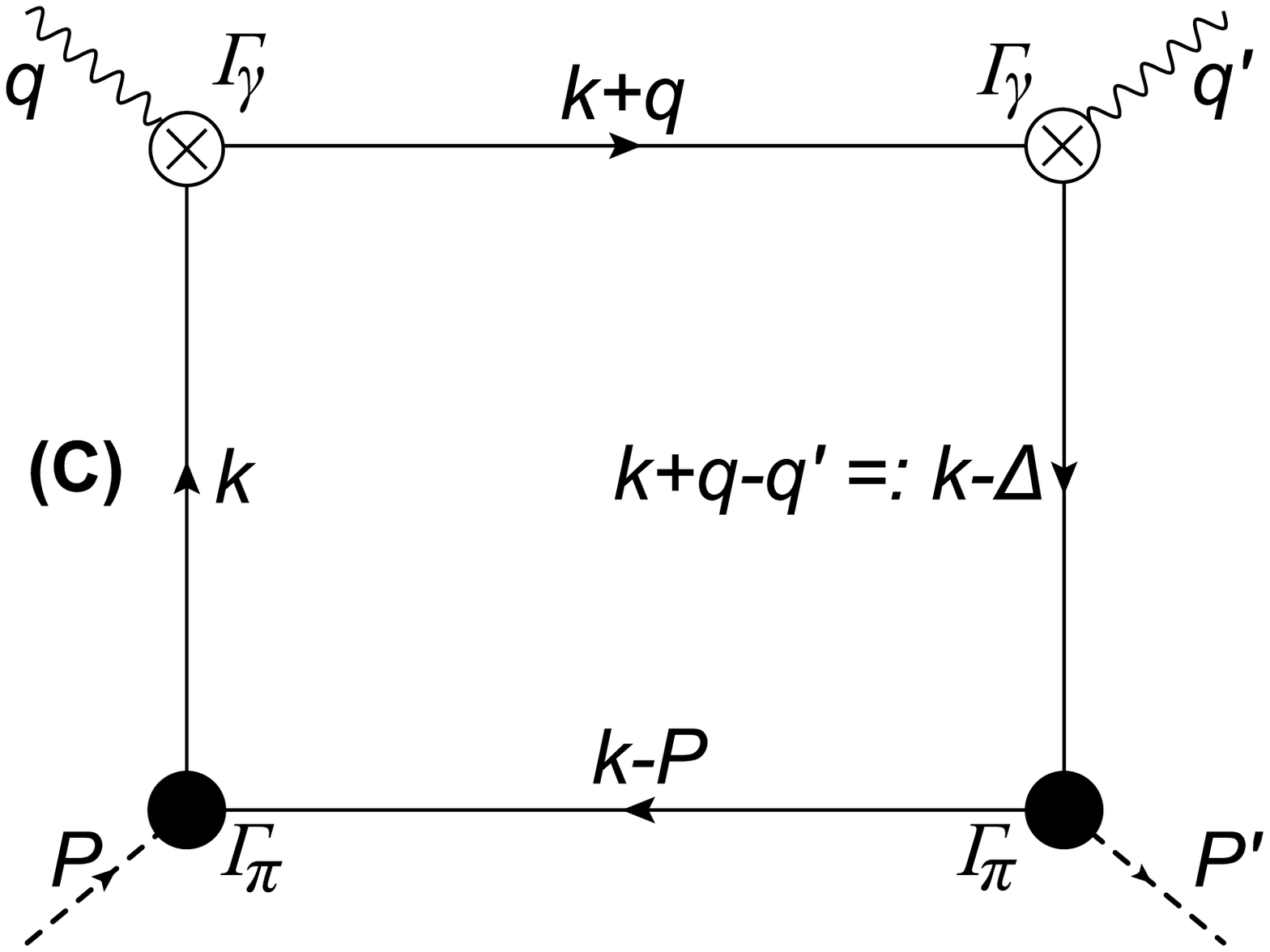}
\caption{\label{figCompton} Amplitude-(1) for virtual-photon--pion Compton scattering in RL truncation is obtained from the sum $(A)+(B)-(C)$.  The ``dots'' in (A) and (B) indicate summation of infinitely many ladder-like rungs.
The other two amplitudes are obtained as follows: (2) -- switch vertices to which $q$ and $q^\prime$ are attached; and (3) -- switch vertex insertions associated with $q^\prime$ and $P^\prime$.
The lines and vertices mean the following:
\emph{dashed line} -- pion;
\emph{undulating line} -- photon;
\emph{spring} -- interaction-gluon in the RL kernels;
\emph{solid line} -- dressed-quark;
\emph{open-cross circle} -- dressed--quark-photon vertex;
\emph{filled circle} -- pion Bethe-Salpeter amplitude.  Each of the last three is computed in RL truncation. }
\end{minipage} 
\end{figure}

\begin{figure}[h]
\begin{minipage}{17.5pc}
\includegraphics[width=17.5pc]{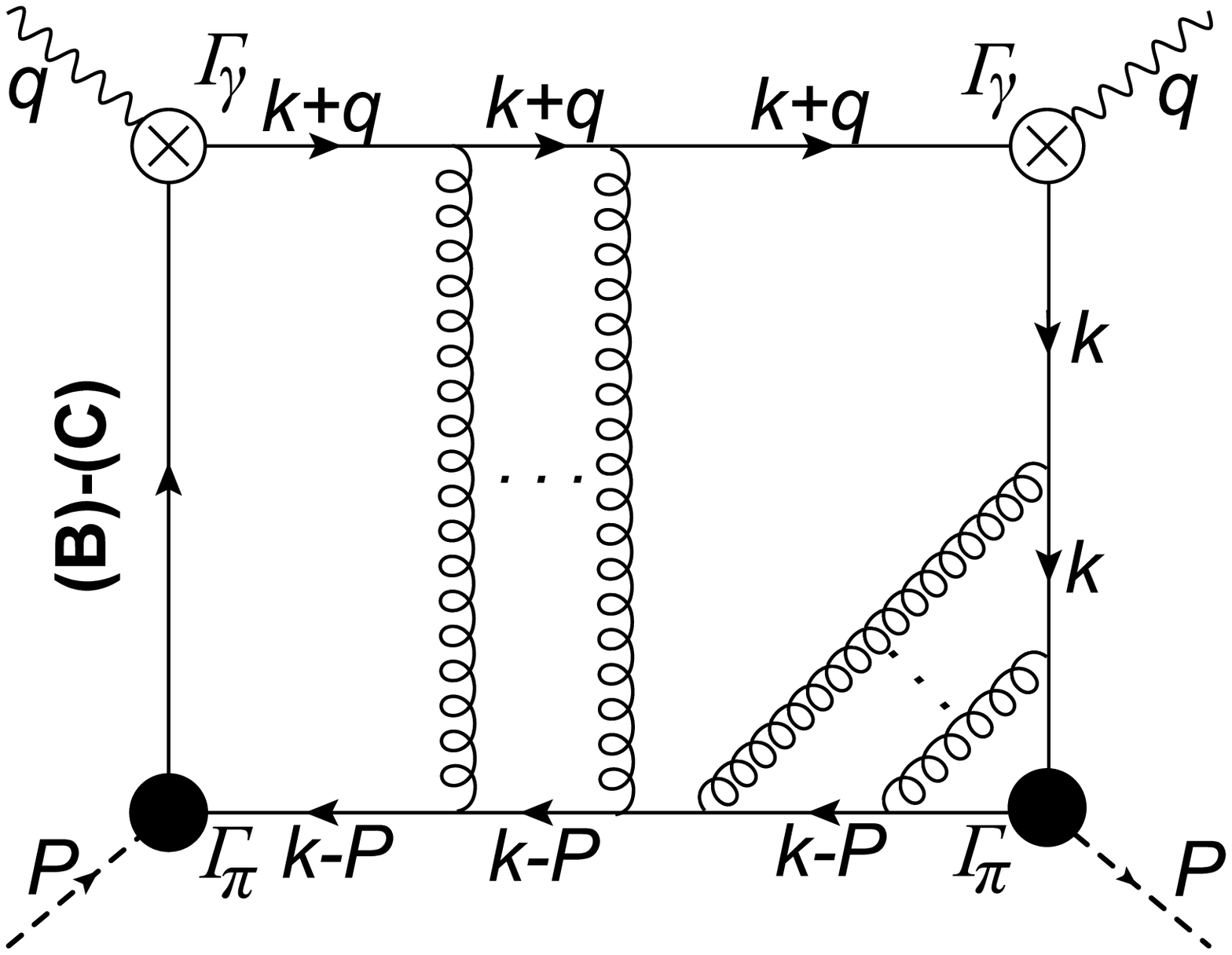}
\caption{\label{figComptonBCF} Forward limit of the combination $(B)-(C)$ in Fig.\,\ref{figCompton}.  The figure also exposes the internal structure of the pion's Bethe-Salpeter amplitude obtained in RL-truncation.
In the Bjorken limit, the initial/final-state interactions involve very soft gluons and hence, in combination with the ladder resummation, produce a contribution that is of the same order as Diagram-(A) in Fig.\,\ref{figCompton}.}
\end{minipage}\hspace{2pc}%
\begin{minipage}{17.5pc}
\includegraphics[width=17.5pc]{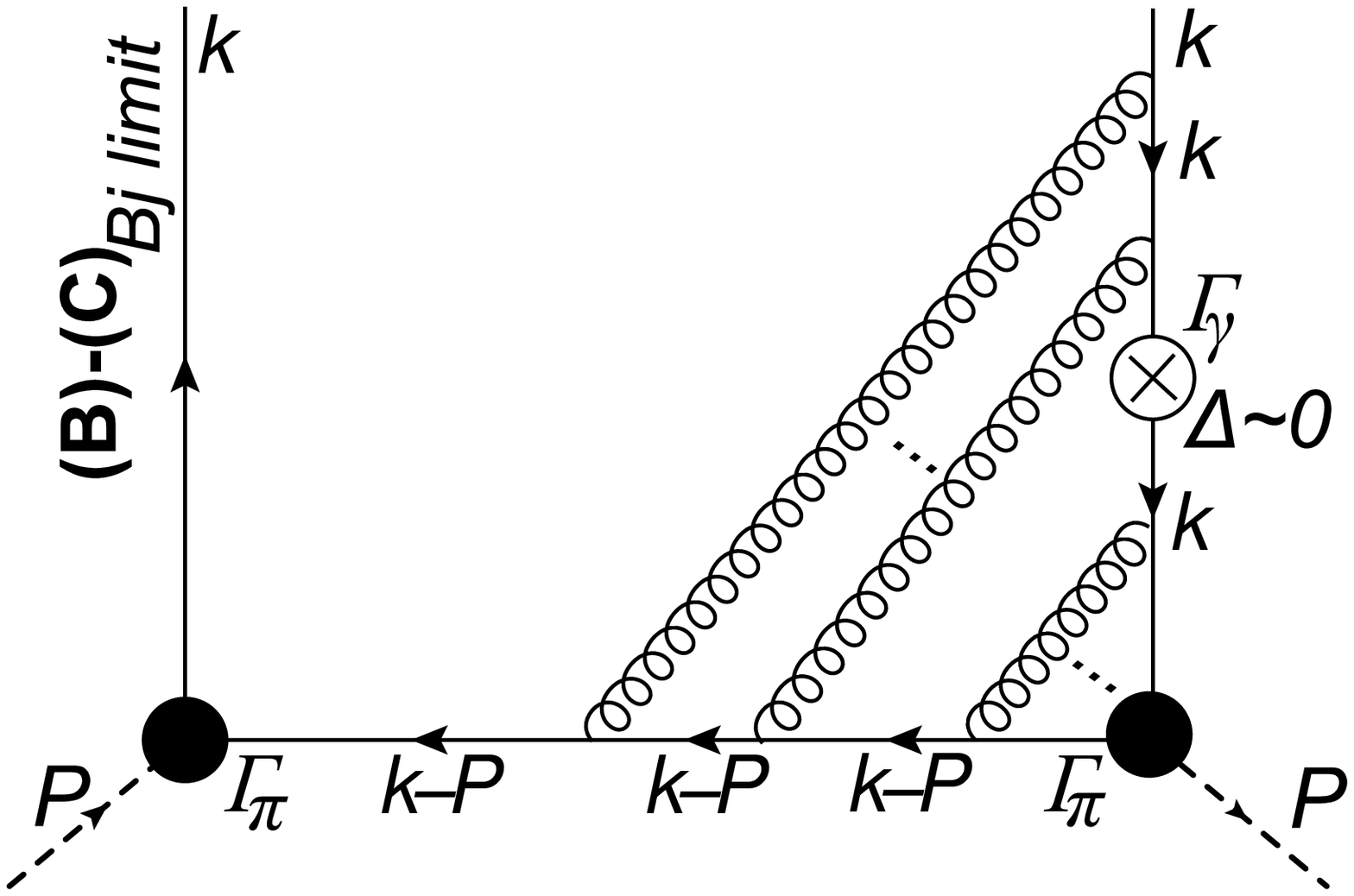}
\caption{\label{figComptonBCFR} Imaginary part of the left panel in the Bjorken limit: the vertex insertion can appear between any two interaction lines.  The compound vertex on the right is readily simplified using the RL Bethe-Salpeter equation.}
\end{minipage} 
\end{figure}

Diagram-(A) in Fig.\,\ref{figCompton} represents the textbook \emph{handbag} contribution to virtual Compton scattering and, 
in the forward and Bjorken limits, yields the following piece of the structure function:
\begin{equation}
\label{qvHklx}
q_{A}^\pi(x) = N_c{\rm tr} \! \int_{dk} \,
\delta_n^{x}(k_\eta)\, n\cdot \gamma \,{\cal H}_\pi(P,k)\,,
\end{equation}
which is typically the only contribution retained in computations of the pion's quark distribution function; e.g., Refs.\,\cite{Shigetani:1993dx,Davidson:1994uv,Bentz:1999gx,Dorokhov:2000gu,Hecht:2000xa,Nguyen:2011jy}. In RL truncation, ${\cal H}_\pi(P,k)$ is an infinite sum of ladder-like rungs, as illustrated in Fig.\,\ref{figCompton}, so that one may write \cite{Nguyen:2011jy}
\begin{equation}
q_{A}^\pi(x)  = N_c  {\rm tr} \! \int_{dk} \!  i\Gamma_\pi(k_\eta,-P) \,S(k_\eta)\, \Gamma^n(k;x) \, S(k_\eta)\, i\Gamma_\pi(k_{\bar\eta},P)\, S(k_{\bar\eta})\,,
\label{Eucl_pdf_LR_Ward}
\end{equation}
wherein $\Gamma^n(k;x)$ is a generalisation of the quark-photon vertex, describing a dressed-quark scattering from a zero momentum photon.  It satisfies a RL Bethe-Salpeter equation with inhomogeneity $i n\cdot\gamma \,\delta_n^{x}(k_\eta)$ \cite{Nguyen:2011jy}.  Equation~\eqref{Eucl_pdf_LR_Ward} ensures \mbox{$ \int_0^1 dx \,  q_A^\pi(x) = 1$} because $ \int dx \, \Gamma^n(\ell;x)$ gives the Ward-identity vertex and the Bethe-Salpeter amplitude is canonically normalised.  The minimal \emph{Ansatz} sufficient to preserve these qualities is $\Gamma^n(k;x) =  \delta_n^{x}(k_\eta)\, \partial_{k_\eta} S^{-1}(k_\eta)$,
in which case one has
%
\begin{equation}
q_{A}^\pi(x) = N_c {\rm tr} \! \int_{dk}
\delta_n^{x}(k_\eta)\Gamma_\pi(k_\eta,-P) \partial_{k_\eta}S(k_\eta)\Gamma_\pi(k_{\bar\eta},P)\, S(k_{\bar\eta})\,.
\label{qAPDF}
\end{equation}
As discussed in Ref.\,\cite{Mezrag:2014jka}, the forward limit for the GPD defined by Eq.\,\eqref{eq:TriangleDiagrams},  $H_{\pi}^{\rm v}(x,0,0)$, corresponds exactly to the pion PDF contribution given by  \eqref{qAPDF}, $q_A^\pi(x)$. However, as proven in Ref.\,\cite{Chang:2014lva}, that impulse approximation is incomplete because it omits a fraction of the contributions from gluons which bind dressed-quarks into the pion. These contributions appear depicted in the diagram of Fig.\,\ref{figComptonBCF}.
As has been lengthly explained in Ref.\,\cite{Chang:2014lva}, the contribution depicted in Fig.\,\ref{figComptonBCF}, which has hitherto been overlooked, is of the same order as that from Diagram-(A), leading to the result of \eqref{qAPDF}. 
Consider now the imaginary part of Fig.\,\ref{figComptonBCF} in the Bjorken limit, which produces the leading contribution illustrated in the Fig.\,\ref{figComptonBCFR}: the vertex insertion can appear between any pair of interaction lines.  Using the recursive structure of the ladder Bethe-Salpeter kernel and the Ward identity, which entails that inserting a zero-momentum vector-probe into a propagator line is equivalent to differentiation of the propagator, then the compound vertex on the right side of Fig.\,\ref{figComptonBCFR} is readily seen to correspond to differentiation of the Bethe-Salpeter amplitude itself with respect to $k_\eta$.  One thus arrives at the following contribution from $(B)-(C)$ to the pion's quark distribution function:
\begin{equation}
q_{BC}^\pi(x) = N_c {\rm tr}\! \int_{dk}\!
\delta_n^x(k_\eta)\partial_{k_\eta} \Gamma_\pi(k_\eta,-P)  S(k_\eta)\Gamma_\pi(k_{\bar\eta},P)\, S(k_{\bar\eta})\,.
\label{qBCPDF}
\end{equation}
This expression is nonzero in general and only vanishes when the pion's Bethe-Salpeter amplitude is independent of relative momentum, so that one is left with
\begin{equation}
q_V^\pi(x) = q_A^\pi(x) + q_{BC}^\pi(x) \, ,
\end{equation}
for the total quark distribution in RL truncation. 

Therefore, since Eq.\,\eqref{eq:TriangleDiagrams} is incomplete for $q_V^\pi(x)$ it is necessarily also inadequate for computation of $H_\pi^{\rm v}(x,\xi,t)$.  Importantly, we have found that the flaw is expressed more forcefully as ${\mathpzc t}:=\Delta_\perp^2$ grows: one can obtain $H_\pi^{\rm v}(x,0,-{\mathpzc t})<0$, which is physically impossible, as explained in connection with Eqs.\,\eqref{overlapGPD}, \eqref{HPositive}.  

The  amendment to Eq.\,\eqref{eq:TriangleDiagrams} is to be guided by the precise knowledge of the form for the correction  $q_{BC}^\pi(x)$, as the former should be non-forward latter's extension in the case of the pion valence-quark GPD. However, it is still being sought and below, for illustrative purposes, we content ourselves with the following ansatz to aument Eq.\,\eqref{eq:TriangleDiagrams} [$d_\pm^n = n\cdot \partial_{\ell_\pm^{\rm R}}$]
\begin{align}
\nonumber  H_\pi^{\rm C}&(x,0,-{\mathpzc t}) =
\frac{1}{2} N_c {\rm tr}\!\!\!
\int_{d\ell}\,\delta_n^{xP}(\ell)\left[
d_+^n\Gamma_\pi(\ell_+^{\rm R};-P_+)
S(\ell_P)\Gamma_\pi(\ell_-^{\rm R};P_-) \right.\\
&
\left.  \times S(\ell_-) + \Gamma_\pi(\ell_+^{\rm R};-P_+)S(\ell_P)
d_-^n\Gamma_\pi(\ell_-^{\rm R};P_-) S(\ell_-)
\right]\,. \label{HCorrection}
\end{align}
This \emph{Ansatz} extends the handbag diagram correction $q_{BC}^\pi(x)$ identified in Ref.\,\cite{Chang:2014lva} to ${\mathpzc t}>0$; and, in connection with the valence dressed-quark GPD, it can be expressed via a Radon amplitude $F^{\rm C}(\alpha,\beta,t)$ which preserves the good features of the kindred amplitude produced by Eq.\,\eqref{eq:TriangleDiagrams}. 

We will now focus primarily on $\xi=0$, be guided by Eq.\,\eqref{eq:TriangleDiagrams}, and mention and ameliorate its failings where appropriate, drawing on the insights gained from the example in Sect.\,3.  Notably, the defects of Eq.\,\eqref{eq:TriangleDiagrams} are typically overlooked in extant continuum computations of the pion's GPD \cite{Tiburzi:2002tq,Broniowski:2003rp,Ji:2006ea,Broniowski:2007si,Frederico:2009fk}): Refs.\,\cite{Broniowski:2003rp,Broniowski:2007si} deliver a form for $q(x,|\vec{b}_\perp|)$ that is not positive definite.

In order to gain novel insights into pion structure, we use the following algebraic forms for the dressed-quark and pion elements in Eq.\,\eqref{eq:TriangleDiagrams} $[\Delta_M(\ell^2)=1/(\ell^2+M^2)]$ \cite{Chang:2013pq}:
\begin{subequations}
\label{NakanishiASY}
\begin{eqnarray}
\label{eq:sim1}
S(\ell) & = &[-i\gamma\cdot \ell+M]\Delta_M(\ell^2)\,,\\
\rho_\nu(z) & = & \frac{1}{\sqrt{\pi}}\frac{\Gamma(v+3/2)}{\Gamma(\nu+1)}(1-z^2)^\nu\,,\\
\label{eq:sim2}
%
%
\mathpzc{n}_\pi \Gamma_\pi(\ell^{\rm R}_\pm;\pm P) & = & i\gamma_5\int^1_{-1}dz\, \rho_\nu(z) \, \hat\Delta^\nu_M(\ell^2_{z\pm})\,,
\label{NoF}
\end{eqnarray}
\end{subequations}
where $M$ is a dressed-quark mass-scale; $\hat\Delta_M(\ell^2) = M^2 \Delta_M(\ell^2)$; $\ell_{z\pm}=\ell^{\rm R}_\pm + (z \pm 1) P/2$ and we work in the chiral limit ($P^2=0=\hat m$, where $\hat m$ is the current-quark mass); and $\mathpzc{n}_\pi$ is the Bethe-Salpeter amplitude's canonical normalisation constant.  Owing to the simplicity of Eqs.\,\eqref{NakanishiASY}, one can reasonably employ $\Gamma_\mu(\ell_+,\ell_-) = \gamma_\mu P_T({\mathpzc t} = \Delta_\perp^2)$, where $P_T({\mathpzc t})$ is the vertex dressing function described in Eqs.\,(24)--(28) of Ref.\,\cite{Roberts:2011wy}. 

\begin{figure}[h]
\includegraphics[width=18.5pc]{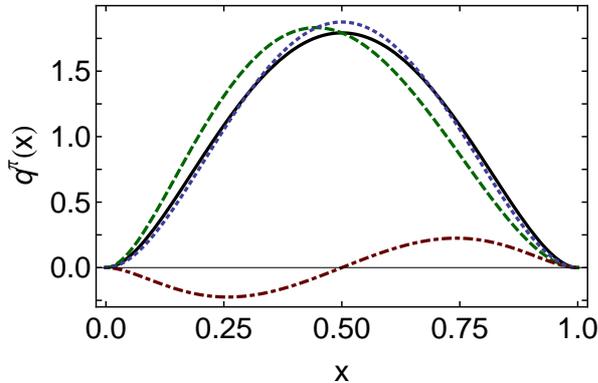}\hspace{2pc}%
\begin{minipage}[b]{17.5pc}\caption{\label{PDFplot} Pion dressed-quark distribution function in rainbow-ladder truncation: \emph{solid} -- complete result, suming Eqs.\,\eqref{qAresult}; \emph{dashed} -- handbag contribution usually retained, Eq.\,\eqref{qAPDF}; and \emph{dot-dashed} -- amendment described in connection with Eq.\,\eqref{qBCPDF}.  An interesting comparison is provided by the \emph{dotted curve}: $q_{2}(x)=30 x^2(1-x)^2$.  To the eye, it is barely distinguishable from our complete result; and the mean value of the absolute relative difference between the curves is just 9\%.}
\end{minipage}
\end{figure}

First, working with the input specified in connection with Eqs.\,\eqref{NakanishiASY}, one can obtain the two contributions for the valence-quark parton distribution, $q_V^\pi(x)$, from Eqs.\,(\ref{qAPDF},\ref{qBCPDF}),
\begin{subequations}
\begin{eqnarray}
\nonumber 
q_A^\pi(x) &=&
\frac{72}{25}
\left[x^3 (x [-2 (x-4) x-15]+30) \ln (x)+ (2 x^2+3) \right.\\
&&\left. \times (x-1)^4 \ln (1-x)+x [x (x [2 x-5]-15)-3] (x-1)\right]\,,
\label{qAresult} \\
\nonumber
q_{BC}^\pi(x) &=& \frac{72}{25}
\left[ x^3 (2 x ([x-3] x+5)-15) \ln (x) -(2 x^3+4 x+9) \right.\\
&& \left. \times (x-1)^3 \ln (1-x) -x (2 x-1) ([x-1] x-9) (x-1)\right]\,;
\end{eqnarray}
\end{subequations}
and display them in Fig.~\ref{PDFplot}. Only after summing the two contributions, $q_A^\pi(x)$ and $q_{BC}^\pi(x)$,  considering  hence the leading contributions from all the diagrams in Fig.\,\ref{figCompton}, the valence-quark parton distribution is symmetric under the exchange $x \longleftrightarrow 1-x$: {\it viz.} the dressed-quark and -antiquark each carry half the pion's momentum.

We computed then, with the same input specified in connection with Eqs.\,\eqref{NakanishiASY}, the triangle diagram result from Eq.\,\eqref{eq:TriangleDiagrams} and the correction given by Eq.\,\eqref{HCorrection}, both to obtain all the leading contributions to the pion's valence dressed-quark GPD, $H_\pi^{\rm v}(x,\xi,t)$.  As detailed elsewhere \cite{Mezrag:2014tva}, that task for triangle diagram was completed by deriving an expression for the Mellin moments of $H_\pi^{\rm v}(x,\xi,t)$ from Eq.\,\eqref{eq:TriangleDiagrams} and introducing five Feynman parameters ($x, y, u, v, w$), defined in the domain $[0, 1]$, and two convolution parameters $z, z^\prime\in [-1, +1]$, so that the momentum integrals could be computed analytically.  Inspecting the result, one can then determine Radon amplitudes for use in Eq.\,\eqref{radon} that are consistent with those moments: the amplitudes vanish outside $\Omega=\{(\alpha,\beta): |\alpha| + |\beta|\leq 1\}$, $F(\alpha,\beta,t)$ is an even function of $\beta$ and $G(\alpha,\beta,t)$ is odd.  Eq.\,\eqref{radon} then entails that $H_\pi^{\rm v}(x,\xi,t)$ complies with the known constraints on polynomiality in $\xi$, vanishes outside $x\in[-\xi,1]$ and is continuous at $x=\xi$. The same recipe is to be applied to Eq.\,\eqref{HCorrection} in order to determine the correction's Radon amplitude $F^{\rm C}(\alpha,\beta,t)$. 
It worth to highlight, once again, that the results from Eqs.\,\eqref{eq:TriangleDiagrams} and \eqref{HCorrection}, in the appropriate forward limit, exactly correspond with those from Eqs.\,\eqref{qAPDF} and \eqref{qBCPDF}, respectively, shown by Eqs.\,\eqref{qAresult}.

We note now that when considering the comprehensive GPD defined by Eq.\,\eqref{eq-def-GPD-H-spinless-target}, one may write with complete generality:
\begin{subequations}
\label{Hgeneral}
\begin{align}
H_\pi(x,0,-{\mathpzc t}) &= H_\pi(x,0,0) {\cal N}({\mathpzc t})
{\cal C}_\pi(x,{\mathpzc t}) F_\pi({\mathpzc t}) \,, \\
1 & =  {\cal N}({\mathpzc t}) \int_{-1}^1dx\, H_\pi(x,0,0)
{\cal C}_\pi(x,{\mathpzc t})\,,
\end{align}
\end{subequations}
so that all $(x,{\mathpzc t})$ correlations in $H_\pi$ are expressed by ${\cal C}_\pi(x,{\mathpzc t})$, which is necessarily non-unity in any physical system \cite{Burkardt:2002hr}.  It is plain from Eq.\,\eqref{radon} that only $F(\alpha,\beta,t)$ contributes when $\xi=0$.

\begin{figure}[h]
\begin{minipage}{17.5pc}
\includegraphics[width=17.5pc]{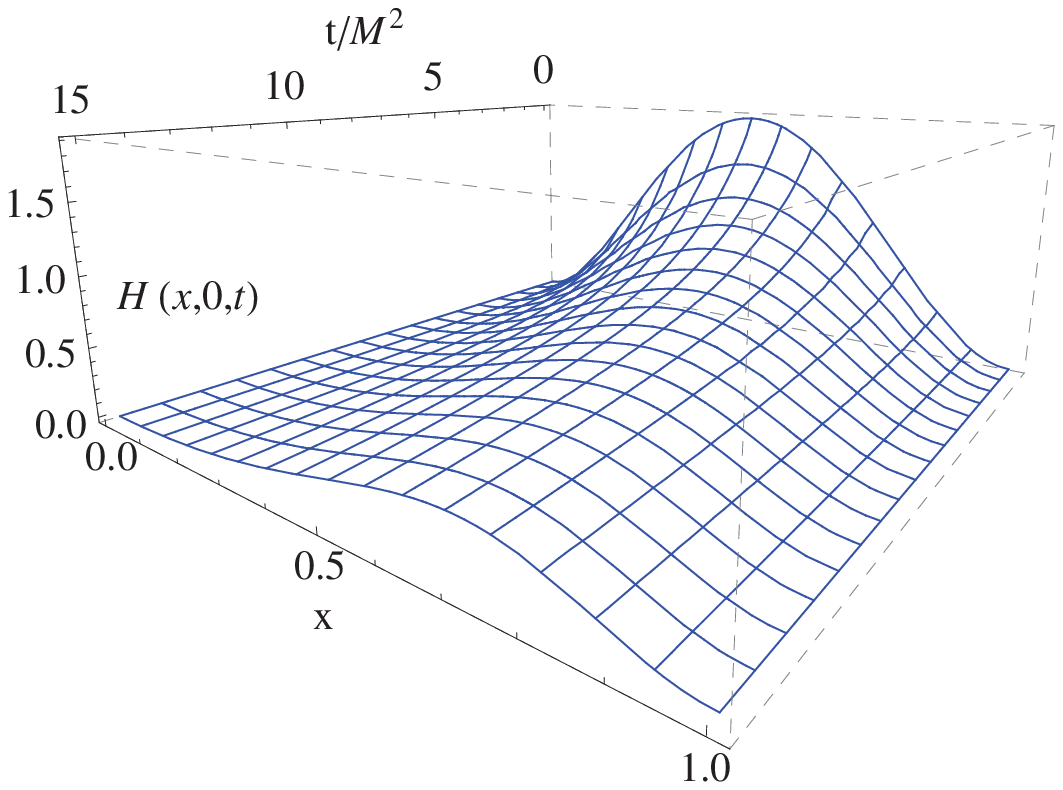}
\caption{\label{figHx0tz0z2} Pion valence dressed-quark GPD, $H_\pi^{\rm v}(x,0,-{\mathpzc t})$, defined by Eqs.\,\eqref{Hgeneral}, \eqref{Cfinal}, obtained at the model scale, $\zeta_H=0.51\,$GeV;}
\end{minipage}\hspace{2pc}%
\begin{minipage}{17.5pc}
\includegraphics[width=17.5pc]{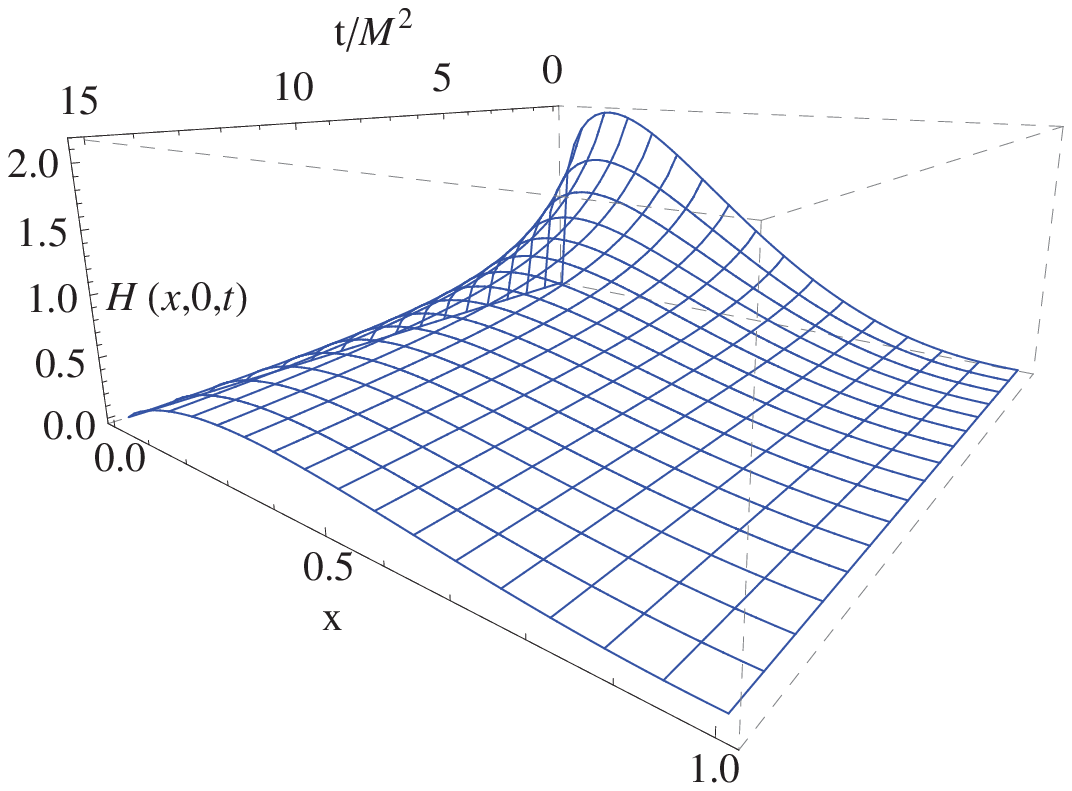}
\caption{The same GPD evolved to $\zeta_2=2\,$GeV using leading-order DGLAP equations.}
\end{minipage} 
\end{figure}

Summing the contributions from Eqs.\,\eqref{eq:TriangleDiagrams}, \eqref{HCorrection}, the net result has the form
\begin{subequations}
\begin{align}
\label{Fdefine}
F(\alpha,\beta,-{\mathpzc t})
&=\phi(\alpha,\beta,{\mathpzc t})^2[F_{\rm S}(\alpha,\beta)
+ {\mathpzc t} V(\alpha,\beta)\phi(\alpha,\beta,{\mathpzc t})] ,\\
\phi(\alpha,\beta,{\mathpzc t}) & =
1/[1+(t/[4M^2])(1-\alpha+\beta)(1-\alpha-\beta)]\,,
\end{align}
\end{subequations}
where the $F_{\rm S}$ component yields $H_\pi^{\rm v}(x,0,0)=q_V^\pi(x)$ in Ref.\,\cite{Chang:2014lva} and that with $V$ is responsible for all violations of Eq.\,\eqref{HPositive}.

\begin{figure}[h]
\includegraphics[width=17.5pc]{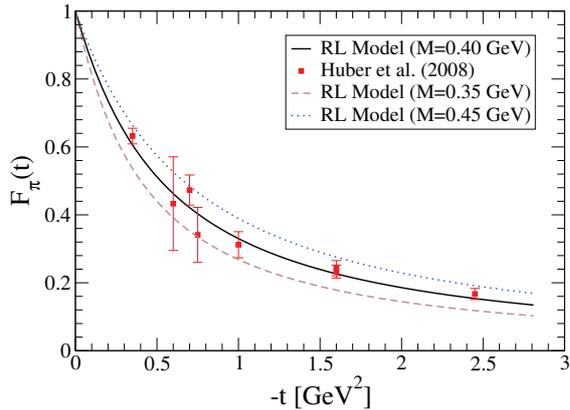}\hspace{2pc}%
\begin{minipage}[b]{17.5pc}\caption{\label{figFpit} Pion electromagnetic form factor obtained from $H_\pi^{\rm v}(x,0,-{\mathpzc t})$, defined by Eqs.\,\eqref{Hgeneral}, \eqref{Cfinal}, which is deliberately consistent with the result determined using Eqs.\,\eqref{eq:TriangleDiagrams}, \eqref{NakanishiASY} and associated definitions.  The data are described in Ref.\,\protect\cite{Huber:2008id}. The most favourable comparison is obtained with $M=0.40\,$GeV in Eqs.\,\eqref{NakanishiASY} and the band shows results with $M=0.40\pm 0.05\,$GeV.}
\end{minipage}
\end{figure}

Acting upon these observations, we define an ameliorated RL approximation to the pion's valence dressed-quark GPD as the function obtained by: setting $V\equiv 0$ in Eq.\,\eqref{Fdefine}; and, for added simplicity, working with $\phi(\alpha,\beta=0,{\mathpzc t})$ whilst keeping the form of $F_\pi({\mathpzc t})$ computed directly from Eq.\,\eqref{eq:TriangleDiagrams}.  Namely, via Eq.\,\eqref{radon}, our valence-quark GPD is given by Eq.\,\eqref{Hgeneral} with
\begin{equation}
\label{Cfinal}
{\cal C}(x,{\mathpzc t}) = 1/[ 1 +({\mathpzc t}/[4M^2]) (1-x)^2 ]^2 .
\end{equation}

Our computed GPD is depicted in the upper panel of Fig.\ref{figHx0tz0z2}.  Notably, the properties described in association with Eqs.\,\eqref{HPositive}--\eqref{eqCz} are evident, and this GPD naturally reproduces the pion valence dressed-quark distribution function obtained in Ref.\,\cite{Chang:2014lva}.

The pion form factor associated with our GPD is drawn in Fig.\,\ref{figFpit}.  A fit to the result is provided by
\begin{equation}
\label{eqFpit}
F_\pi({\mathpzc t}=M^2 z)= \frac{1+0.16 z}{1+0.44 z + 0.060 z^2 + 0.00033 z^3}\,.
\end{equation}
At large-$\mathpzc t$ it behaves as $1/{\mathpzc t}^2$, whereas the correct power-law dependence is $1/{\mathpzc t}$ \cite{Farrar:1979aw,Efremov:1979qk,Lepage:1980fj}.  The power-law is wrong because Eq.\eqref{NoF} omits terms that have been described as representing the pion's pseudovector components \cite{Maris:1998hc}, which are necessarily nonzero in a complete picture of the physical pion \cite{Maris:1997hd,Qin:2014vya}.  Notwithstanding that, it is valid and useful to compare the prediction with contemporary data and thereby determine a sensible value for our model's dressed-quark mass-scale: the best comparison is obtained with $M=0.4\,$GeV.  Notably, this scale is typical of the dressed-quark mass function in QCD \cite{Bhagwat:2003vw,Bowman:2005vx,Bhagwat:2006tu}.

\begin{figure}[h]
\begin{minipage}{17.5pc}
\includegraphics[width=17.5pc]{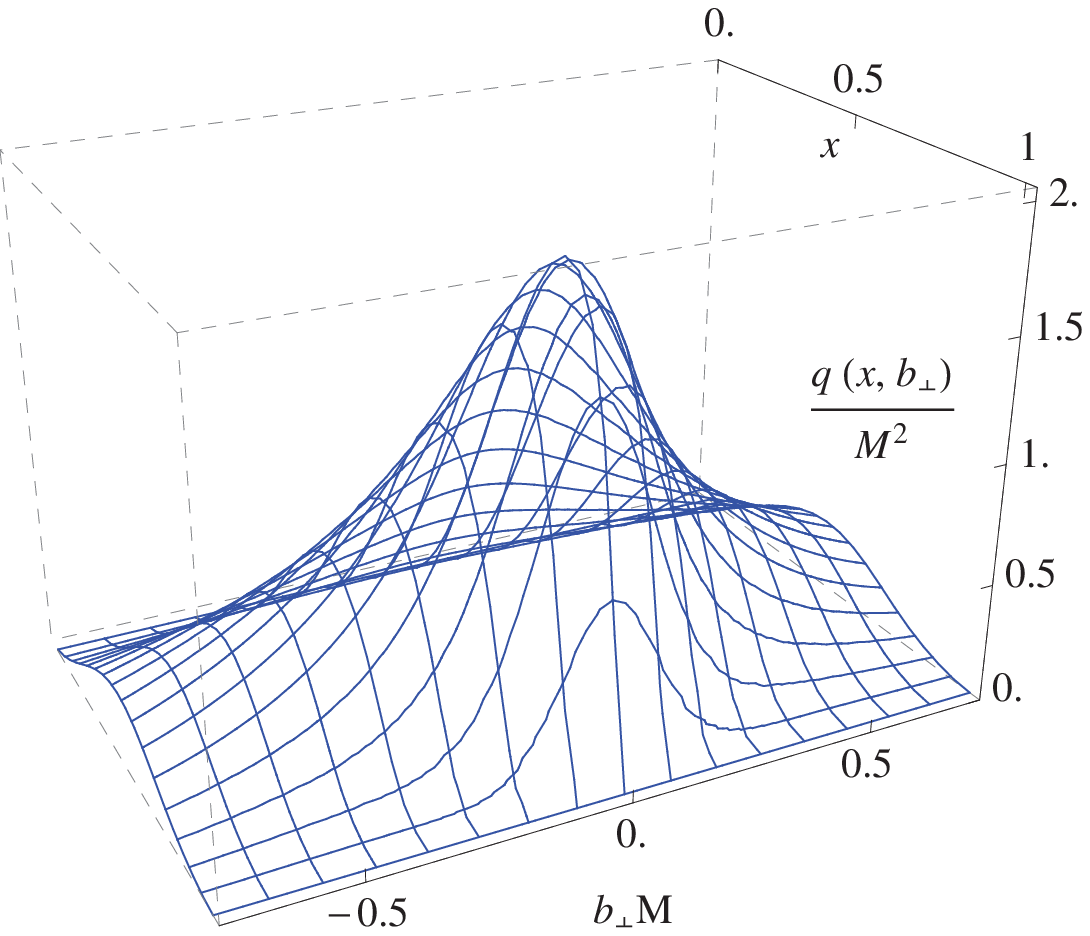}
\caption{\label{figqbx2} Pion's valence dressed-quark GPD in impact parameter space, $q_\pi^{\rm v}(x,|\vec{b}_\perp|;\zeta)$. Result obtained from $H_\pi^{\rm v}(x,0,-{\mathpzc t};\zeta_H)$ in the top panel of Fig.\,\ref{figHx0tz0z2} using Eq.\,\eqref{defqxb}.}
\end{minipage}\hspace{2pc}%
\begin{minipage}{17.5pc}
\includegraphics[width=17.5pc]{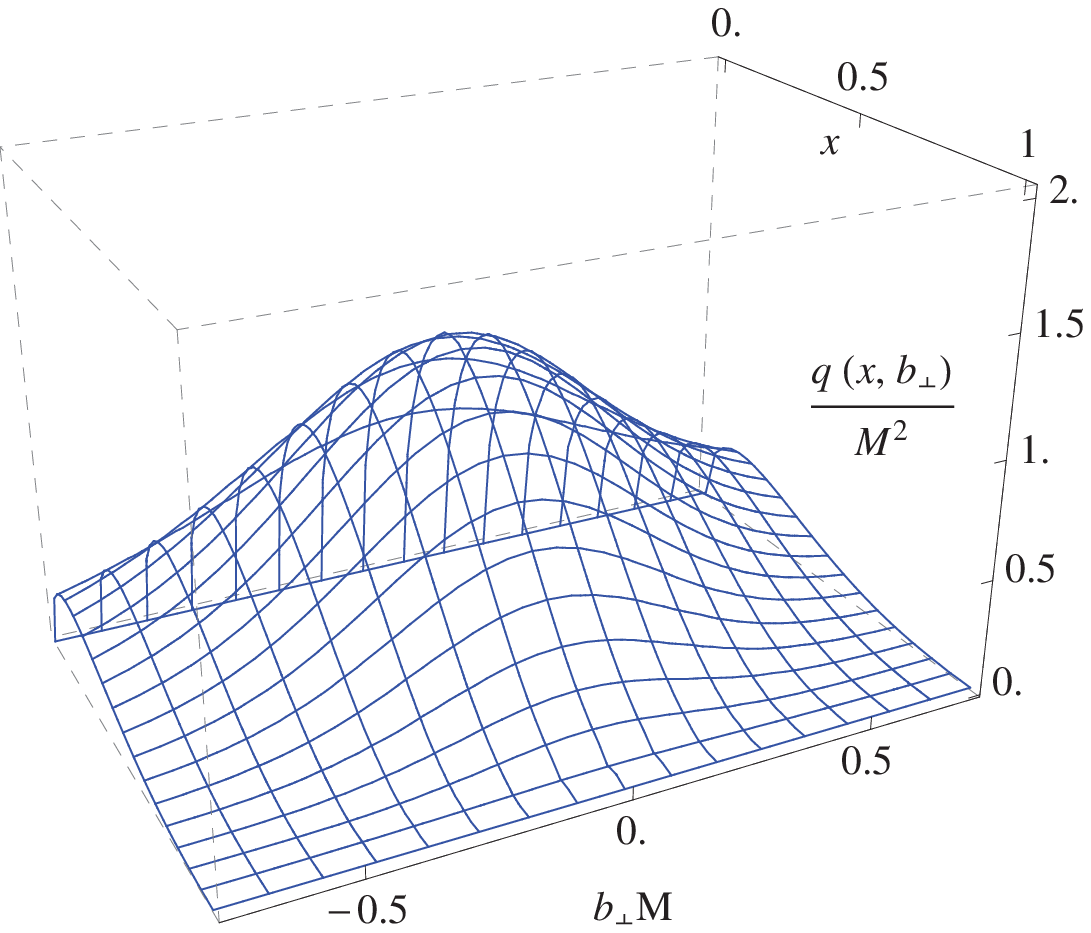}
\caption{\label{label} Analogous result associated with $H_\pi^{\rm v}(x,0,-{\mathpzc t};\zeta_2)$ in the bottom panel of Fig.\,\ref{figHx0tz0z2}, see Sect.\,5.  [N.B.\,$1/M \approx 0.5\,$fm, so $ b_\perp M =0.5$ corresponds to $b_\perp \approx 0.25\,$fm and $ q_\pi^{\rm v}(x,|\vec{b}_\perp|;\zeta)/M^2 =1$ means $q_\pi^{\rm v}(x,|\vec{b}_\perp|;\zeta)\approx 4\,$fm$^{-2}$.}
\end{minipage} 
\end{figure}

The IPD GPD derived from $H_\pi^{\rm v}(x,0,-{\mathpzc t})$ in the upper panel of Fig.\,\ref{figHx0tz0z2} is depicted in the upper panel of Fig.\,\ref{figqbx2}.  The global maximum in this valence distribution is located at $(x=0.76,|\vec{b}_\perp|=0)$ and, plainly, the probability to find a dressed-quark is strongly localised around this maximum.  Naturally, for this valence dressed-quark distribution ($d^2 |\vec{b}_\perp| = 2\pi \,d |\vec{b}_\perp|\, |\vec{b}_\perp|$):
\begin{equation}
\begin{array}{c}
\int_{-1}^1dx\!\int_0^\infty d^2 |\vec{b}_\perp|\,x \,q_\pi^{\rm v}(x,|\vec{b}_\perp|)
=\frac{1}{2}.
\end{array}
\end{equation}

\smallskip


\section{Evolution of the GPD} 
As explained elsewhere \cite{Chang:2014lva}, our framework yields a valence-quark GPD that may be associated with an hadronic scale $\zeta_{\rm H}=0.51\,$GeV.  It is worth outlining how the features of this distribution evolve to higher scales.  Given that we have used $\xi=0$, that is readily accomplished by using the DGLAP evolution equations to determine the evolved $x$-profile at each value of ${\mathpzc t}$.  Our aim is to provide a qualitative illustration so, unlike Ref.\,\cite{Chang:2014lva}, we do not augment the valence distribution via the inclusion of gluon or sea-quark contributions.  If desired, one could mask the impact of this omission by focusing on the behaviour of $x H_\pi^{\rm v}(x,0,t)$ and $xq_\pi^{\rm v}(x,|\vec{b}_\perp|)$.

Beginning with the valence dressed-quark GPD in the upper panel of Fig.\,\ref{figHx0tz0z2}, we employed leading-order evolution to obtain $H_\pi^{\rm v}(x,0,-{\mathpzc t};\zeta_2=2\,{\rm GeV})$.\footnote{Any significant differences generated by next-to-leading-order evolution are masked by a 25\% increase in $\zeta_{\rm H}$ \cite{Gluck:1999xe} and hence are immaterial herein.}  The result is depicted in the lower panel of Fig.\,\ref{figHx0tz0z2}.  Evidently, evolution, which adds glue and sea-quarks to the system by exposing this substructure within the dressed-quark, sharpens the peak associated with the global maximum at ${\mathpzc t}=0$ and shifts its location toward $x=0$.  The maximum value at each ${\mathpzc t}\neq 0$ is also shifted toward $x=0$; but outside a neighbourhood of ${\mathpzc t}=0$ the profile in $x$ is progressively flattened with increasing ${\mathpzc t}$.  Notwithstanding this, at any finite $\zeta>\zeta_2$, there will be a ${\mathpzc t}_1$ such that $\forall {\mathpzc t}> {\mathpzc t}_1$ a peak, albeit with much suppressed height, may be said to exist in the neighbourhood $x\simeq 1$: ${\mathpzc t}_1$ increases with $\zeta$.

The last observation leads us to consider the conformal limit of QCD, which is recovered on $\tau \simeq 0$, $\tau=\Lambda_{\rm QCD}/\zeta$.  Within this domain, the valence dressed-quark GPD is $H_\pi^{\rm v}(x,0,0;\tau \simeq 0)=\delta(x)$ \cite{Georgi:1951sr,Gross:1974cs,Politzer:1974fr}.  [Fig.\,\ref{figHx0tz0z2} highlights that this limit is reached slowly because evolution is logarithmic in QCD.]  Eqs.\,\eqref{Hgeneral} then entail:
\begin{equation}
\label{Hvtau0}
H_\pi^{\rm v}(x,0,-{\mathpzc t};\tau \simeq 0) = \delta(x) F_{\pi}({\mathpzc t})\,.
\end{equation}
This is a feature of our approximation to the pion's valence dressed-quark GPD.  It is not a property of the pion's complete GPD, $H_\pi(x,0,t;\tau \simeq 0)$, because the valence GPD is a negligible piece of the full GPD on $\tau\simeq 0$.  That may be seen, e.g., by noting that valence-quarks carry none of the pion's momentum within the conformal domain and hence it is invalid therein to represent $F_\pi(t)$ by an impulse (rainbow-ladder) approximation expressed through the triangle diagram of Eq.\,\eqref{eq:TriangleDiagrams}.

Having determined $H_\pi^{\rm v}(x,0,-{\mathpzc t};\zeta_2)$, it is straightforward to obtain $q_\pi^{\rm v}(x,|\vec{b}_\perp|;\zeta_2)$ from Eq.\,\eqref{defqxb}.  The result is depicted in the lower panel of Fig.\,\ref{figqbx2}: apparently, the maximum is shifted toward $x=0$ and compressed in that direction, the peak height is diminished, and the width of the distribution in $|\vec{b}_\perp|$ is increased.

Each one of these evolution-induced changes may be intuitively understood by reasoning as follows.
First consider a limiting case of an active parton with $x\approx 1$.  This parton carries (almost) all the longitudinal momentum of the hadron.  It therefore \emph{defines} the CoTM and hence cannot be far removed from that centre.  The distribution associated with an $x\approx 1$ parton must therefore be tightly localised around $|\vec{b}_\perp|=0$.
On the other hand, consider the case of an active parton with $x$ reduced toward the location of the global maximum.  The remaining partons within the hadron share in defining the CoTM and hence the active parton is not constrained to lie at $|\vec{b}_\perp|=0$.  Plainly, as a parton's value of $x$ diminishes toward the favoured value, it plays less of a role in determining the CoTM and may therefore possess even larger values of $|\vec{b}_\perp|$.

In the current context, recall that evolution exposes the glue and sea-quark content of a dressed-quark: its identity comes to be shared amongst a host of partons, so that the probability of any one parton carrying $x\approx 1$ is much diminished.  It follows that the global maximum in $q(x,|\vec{b}_\perp|;\zeta)$ must move toward $x=0$ with increasing $\zeta$ and, simultaneously, that the distribution is broadened in $|\vec{b}_\perp|$ on the remaining domain of material support.

\begin{figure}[h]
\includegraphics[width=17.5pc]{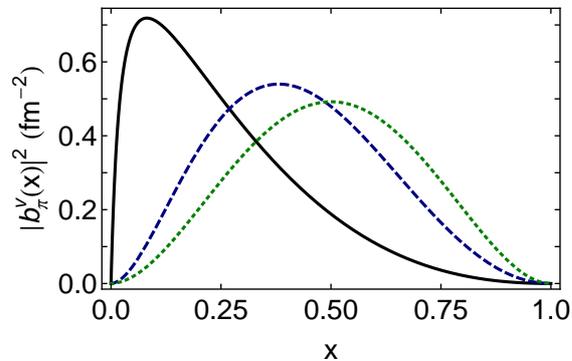}\hspace{2pc}%
\begin{minipage}[b]{17.5pc}\caption{\label{b2x}Distribution of pion's mean-square transverse extent, Eq.\,\eqref{bperp2}: (black) solid curve $\langle|\vec{b}_\perp(x;\zeta_2)|^2 \rangle$; and (blue) dashed curve -- $\langle|\vec{b}_\perp(x;\zeta_H)|^2 \rangle$.  The (green) dotted curve is the result obtained at $\zeta_H$ when the correlation function ${\cal C}_\pi(x,t)$ in Eq.\,\eqref{Cfinal} is neglected: comparison with the dashed curve shows that the product \emph{Ansatz} $H_\pi^{\rm v}(x,0,-{\mathpzc t}) = q_{\rm V}^\pi(x) F_\pi({\mathpzc t})$ is generally unreliable.}
\end{minipage}
\end{figure}

The latter effect is illustrated in Fig.\,\ref{b2x}, which depicts
\begin{equation}
\label{bperp2}
\begin{array}{c}
\langle |\vec{b}_\perp(x;\zeta)|^2 \rangle = \int_0^\infty d^2 |\vec{b}_\perp|
\,  q(x,|\vec{b}_\perp|;\zeta)\,|\vec{b}_\perp|^2\,;
\end{array}
\end{equation}
i.e., the $x$-distribution of the pion's mean-square transverse extent: under evolution, the transverse extent narrows at large-$x$ and broadens at small-$x$.  A little consideration reveals that the measure of the curves in Fig.\,\ref{b2x} is independent of the scale $\zeta$ because evolution is an operation that preserves the area under $H(x,0,t)$ at each $t$.  In fact, using Eqs.\,\eqref{Fpiconnection2}, \eqref{defqxb}, \eqref{bperp2}, one finds
\begin{subequations}
\begin{align}
\langle |\vec{b}_\perp|^2 \rangle & = \int_{-1}^1 \rule{-1ex}{0ex}dx \,
\langle |\vec{b}_\perp(x;\zeta)|^2 \rangle =
\int_0^\infty \rule{-1ex}{0ex}  d|\vec{b}_\perp|\,|\vec{b}_\perp|^3\,{\mathpzc d}_\pi(|\vec{b}_\perp|) \,,\rule{-2ex}{0ex} \\
{\mathpzc d}_\pi(|\vec{b}_\perp|) &= \int_0^\infty d{\Delta} \,{\Delta}\, J_0(|\vec{b}_\perp| {\Delta}) F_\pi({\Delta}^2)\,,
\end{align}
\end{subequations}
and hence, with $F_\pi(t)$ in Eq.\,\eqref{eqFpit}, $\langle |\vec{b}_\perp|^2 \rangle = (0.52\,{\rm fm})^2$.  For the reasons just explained, this is also the value obtained with Eq.\,\eqref{Hvtau0}.  [Note that $F_\pi({\mathpzc t}) = 1/(1 + {\mathpzc t} r_\pi^2/6)$, where $r_\pi$ is the pion's electric charge radius, yields $\langle |\vec{b}_\perp|^2 \rangle = (2/3) r_\pi^2 = (0.55\,{\rm fm})^2$ (empirically \cite{Beringer:1900zz} $r_\pi = 0.672 \pm 0.008\,$fm).]  It is natural that the mean-squared transverse extent of the dressed-quarks within a pion should be commensurate with the length-scale associated with light-quark confinement realised through a violation of reflection positivity [see, e.g., Sect.\,2.2 in Ref.\,\cite{Cloet:2013jya}].


\section{Conclusion and prospects}
We described a calculation of the pion's valence dressed-quark generalised parton distribution (GPD), $H_\pi^{\rm v}(x,\xi,t)$, within the context of a rainbow-ladder (RL) truncation of QCD's Dyson-Schwinger equations.  This framework is useful at an hadronic scale because it provides a description of hadrons via a dressed-quark basis, the accuracy of which in any given channel is knowable \emph{a priori}.  Our analysis focused here on the case of zero skewness ($\xi=0$) but we have also shown in Ref.\,\cite{Mezrag:2014jka} that the symmetry-preserving character of the RL truncation is all one needs to demonstrate a known relationship between $H_\pi^{\rm v}(x,\xi=\pm 1,t)$ and the pion's valence-quark parton distribution amplitude.

Drawing analogy with the pion's valence dressed-quark distribution function, we argued that the impulse-approximation used hitherto to define the pion's valence GPD is generally invalid owing to omission of contributions from the gluons which bind dressed-quarks into the pion.  We used a simple correction [Eq.\eqref{HCorrection}], valid in the neighbourhood of $\xi=0$, $t=0$, in order to identify a practicable improvement to the approximation for $H_\pi^{\rm v}(x,0,t)$.  Expressing the result as the Radon transform of a single amplitude, we were able to isolate and remove those terms which produce unphysical behaviour, such as violations of positivity by the $\xi=0$ GPD.  The resulting, ameliorated Radon amplitude yields a form for $H_\pi^{\rm v}(x,0,t)$ [Eqs.\,\eqref{Hgeneral}, \eqref{Cfinal}] which is consistent with significantly more known constraints than is the result produced by the impulse approximation alone.

The results obtained in this way for $H_\pi^{\rm v}(x,0,t)$, $q_\pi^{\rm v}(x,|\vec{b}_\perp|)$ [Figs.\,\ref{figHx0tz0z2}, \ref{figqbx2}] provide a qualitatively sound picture of the dressed-quark structure of the pion at an hadronic scale.  Using leading-order expressions, we evolved these distributions to a scale $\zeta=2\,$GeV.  All features of the resulting valence quark GPDs may be intuitively understood and hence the distributions should serve as an elementary but reasonable guide in the planning and interpretation of relevant experiments at existing or anticipated \cite{Accardi:2012qut} facilities, which could plausibly involve deeply-virtual Compton scattering on pions in a nucleon's meson cloud.

Notwithstanding the simplicity of the framework employed herein, a merit of the approach is its potential to \emph{compute} features of hadron GPDs on the valence-quark domain and relate them directly to properties of QCD.  This capacity has already been demonstrated in the simpler case of the pion's valence parton distribution function \cite{Chang:2014lva}.  One may begin to realise that potential by using more realistic forms for the dressed-propagators and -vertices that appear in the RL truncation analysis and, perhaps more importantly, uncovering the amendment to impulse approximation which is required in order to extend the validity of the RL truncation to the entire kinematic domain of $\xi$ and $t$.

\section*{Acknowledgments}
We thank
A.~Besse,
I.\,C.~Clo\"et,
D.\,M\"uller,
P.~From\-holz,
C.~Keppel,
P.~Kroll,
J.-Ph.~Lansberg
C.~Lorc\'e,
J.~Segovia,
S.~Schmidt,
P.~Tandy
and
S.~Wallon
for valuable discussions.
CM, LC, HM, CDR and JR-Q are grateful for the chance to participate in the workshop 
``Many Manifestations of Nonperturbative QCD under the Southern Cross'', Ubatuba, S\~ao Paulo, where significant parts of this work were first presented and improvements discussed.
CDR acknowledges support from an \emph{International Fellow Award} from the Helmholtz Association; and
this research was otherwise supported by:
Commissariat \`a l'Energie Atomique;
%
JRA ``Study of Strongly Interacting Matter'' (Grant Agreement no.\,283286, HadronPhysics3)
%
under the 7th E.U. F.P.;
GDR 3034 PH-QCD; 
ANR-12-MONU-0008-01 ``PARTONS'';
University of Adelaide and Australian Research Council through grant no.~FL0992247;
Spanish ministry Research Project FPA2011-23781;
U.S.\ Department of Energy, Office of Science, Office of Nuclear Physics, contract no.~DE-AC02-06CH11357;
and For\-schungs\-zentrum J\"ulich GmbH.


\end{document}